# Suction effects in deep Boom clay block samples

Pierre DELAGE [1], Trung-Tinh LE [1], Anh-Minh TANG [1], Yu-Jun CUI [1], Xiang-Ling LI [2]

[1] École Nationale des Ponts et Chaussées (CERMES, Institut Navier), Paris, France
[2] EURIDICE Group, Mol, Belgium

**Abstract**

Extensive investigations have been and are being carried out on a stiff clay from an underground research laboratory located at Mol (Belgium) called Boom clay, in the context of research into deep nuclear waste disposal. Suction effects in deep Boom clay block samples were investigated through the characterisation of the water retention and of the swelling properties of the clay. The data obtained allowed an estimation of the sample initial suction that was reasonably compatible with the in-situ state of stress at a depth of 223 m. The relationship between suction and stress changes during loading and unloading sequences were also examined by running oedometer tests with suction measurements. *A rather wide range of the ratio $s/\sigma'_v$ (being s the suction and $\sigma'_v$ the effective vertical stress) was obtained (0.61 – 1), different from that proposed by Bishop et al; (1974).* Finally, the effect of suction release under an isotropic stress close to the estimated sample suction was investigated. A slight swelling (1.7%) was observed and further compression provided a satisfactory value of the overconsolidation ratio confirming the suggestions of taking some precautions before putting a swelling sample in contact with water as suggested by Graham et al. (1987). *The various experimental data gathered in this study finally evidenced a relatively good state of conservation of the block sample used.*

# 1. Introduction

Various investigations have been carried out on Boom clay, a stiff clay from Belgium, in the context of research into deep nuclear waste disposal (SAFIR 2, 2001). Investigations are linked to research conducted at the Mol Underground Research Laboratory (URL), excavated in a layer of Boom clay at a depth of 223 m by SCK-CEN, the Belgian organisation for nuclear studies, near the city of Mol.

Most research has been carried out on triaxial specimens that were trimmed from blocks extracted during excavation sequences in the URL. Various questions arose concerning the quality of the specimens and the best procedure to adopt prior to running triaxial tests (Winter and Horseman, 1993; Coll, 2005). Special concern was related to the great depth at which the blocks were extracted, to the corresponding stress release and to the resulting suction (Skempton 1961, Skempton and Sowa 1963, Doran et al. 2000), obviously larger than in common geotechnical engineering practice. The swelling observed in Boom clay specimens when releasing suction by putting them in contact with saturated porous stones in the triaxial cell also provided some concerns because of possible alteration of the natural initial microstructure (Sultan, 1997) with some consequence for the mechanical properties of the clay, and more particularly the overconsolidation ratio.

In this paper, suction effects in deep Boom clay block samples are investigated through the characterisation of the water retention and of the swelling properties of the clay. The relationship between suction and stress changes during loading and unloading sequences are also examined by running oedometer tests with suction measurements. Finally, the effect of suction release under an isotropic stress close to the estimated suction is investigated.

## 2. Material and experimental methods

The study was carried out on intact Boom clay specimens extracted at a depth of 223 m in the URL of Mol, Belgium. The Boom clay formation belongs to the Rupelian geological period in the Tertiary sub-era, which dated from 36 to 30 million years before present. Its thickness is around 102 m in the Mol area (Figure 1) and the layer is gently dipping (± 1°) towards the North-North-East (Mertens et al. 2003). *As commented by Horseman et al. (1987) cited in Burland (1990), the clay is geologically lightly overconsolidated but the yield stress of Boom clay may be larger than the preconsolidation pressure due to mechanisms such as creep and diagenesis, resultanig in an "overconsolidation" ratio $R_{OC} = 2.4$.*

Table 1 presents the mineralogical composition of Boom clay, taken from different works (Decleer *et al.*, 1983; Al-Mukhtar *et al.*, 1996; SAFIR 2, 2001). The clay fraction is dominant (50% to 60% < 2 μm) but some differences exist in the smectite fraction between Al-Mukhtar *et al.* and Decleer *et al.* on one hand (33%) and SAFIR 2 on the other hand (17%). Other differences are observed in the kaolinite and illite fractions. These differences can be due to the natural variability of Boom clay. They show however that some confirmation about the mineralogical composition of Boom clay is still needed. The geotechnical properties of Boom clay are shown in Table 2. *In-situ* water content measurements were made on excavated blocks during the excavation in the URL (connecting gallery, excavated between 23 January and 23 April 2002). Average values between 24.5% and 25.5% were obtained at that time. *Excavated blocks were immediately vacuum packaged in reinforced aluminium foil and thermo-welded. Blocks were stored in a room with temperature varying between 15 and 20°C and an average relative humidity of 45%.* The laboratory water content determination that was made during this study (in 2005) gave smaller values of between 20.2 and 21.6%. The difference in water content between the two measurements is related to some drying that may have occurred during the sample storage between 2002 and 2005. *As compared to the initial suction at the time when the sample was excavated, this drying obviously corresponds to an increase in suction that will be further commented based on the water retention properties of the sample.*

The water retention properties were determined on both rectangular clay samples (30x30x10 mm approximately) and cylindrical oedometer samples (d = 70 mm, h = 20 mm). Starting from initial water contents close to 21%, the rectangular clay samples were dried at controlled values of suction by using the vapour equilibrium method (see for instance Delage *et al.*, 1998) with 5 saline solutions: $CuSO_4$ (*s* = 2.8 MPa); $K_2SO_4$ (*s* = 4.2 MPa); $KNO_3$ (*s* = 8.5 MPa); NaCl (*s* = 37.8 MPa) and $MgCl_2$ (*s* = 152.8 MPa). Along the drying path, three specimens were used at each suction level to determine the water content at equilibrium. Along the wetting path, the three oedometer specimens were smoothly wetted by putting them in contact with humid filter papers and the resulting suction was afterwards measured by using a tensiometer. The volume changes of the oedometer samples were determined with a precision calliper. The volume changes of the rectangular specimens were determined by hydrostatic weighing after having immersed the samples in a non aromatic hydrocarbon liquid called Kerdane.

An oedometer cell equipped with a high suction tensiometer (capable of measuring to 1500 kPa) was used to measure suction changes during oedometer compression, as suggested by Dineen and Burland (1995) (Figure 2). The principle of the high suction tensiometer was first reported by Ridley and Burland (1993). The tensiometer used in this work was built at ENPC-CERMES. As seen in the figure, the tensiometer is located within the base of the oedometer cell in contact with the lower face of the sample, whereas the upper face of the sample is in contact with a porous stone that is fixed to the piston.

Finally, an isotropic compression test was carried out in a high-pressure triaxial cell (see Delage *et al.,* 2000) on a standard triaxial specimen (d = 38 mm, h = 76 mm). The confining pressure and the back-pressure were applied by two GDS volume/pressure controllers. During the test, the sample volume changes were measured by monitoring the volume of water in the cell through the pressure/volume controller used for the confining pressure. As seen further on, this was necessary for i) controlling the volume changes during compression sequence with dry porous stones (i.e. with no water permitted to infiltrate inside the sample) and ii) controlling the swelling when water was allowed to hydrate the sample through the porous stones. In high pressure metal cells, this method appeared to give satisfactory results (Delage et al. 2000).

## 3. Experimental results

*Water retention curve and swelling behaviour*

Figure 3 shows some aspects of the water retention properties of Boom clay presented in terms of changes in water content ($w$) and of degree of saturation ($S_r$) as a function of the logarithm of suction (log $s$). As described previously, starting from initial water contents of 20.2 – 21.6%, three points were obtained along a wetting path (with measured suctions equal to 180, 280 and 600 kPa respectively) and five suctions were used along a drying path (with imposed suctions equal to 2.8, 4.2, 8.5, 37.8 and 152.8 MPa and with three specimens at each suction level). The data obtained along the drying path show a good compatibility between the various points obtained under the same suction, both in terms of water content and degree of saturation.

The $S_r$ - log $s$ plot shows that the two points obtained along the drying path at the two low suctions (2.8 and 4.2 MPa) indicate that the samples remained saturated. Desaturation starts above 4.2 MPa and the degree of saturation at a 8.5 MPa suction is 90%. As shown in the figure, the air entry value of Boom clay can be estimated at approximately 5 MPa. At the highest suction (152.8 MPa), the degree of saturation is equal to 31%.

Along the wetting path, the curve shows that, curiously, the degree of saturation of the oedometer samples decreases below 100%, with values lying between 90 and 100%. Since the volume change measurements carried out by hydrostatic weighing on the samples subjected to suctions of 2.8 and 4.5 MPa showed that the sample remained saturated, it is believed that the initial state of the sample close to $w = 21\%$ was saturated (unfortunately, no determination of the degree of saturation was made in the initial state). Hence, the samples subjected to suctions lower than the initial one that reached higher values of water content (22.5, 29 and 29.5%) should not desaturate. The value of degree of saturation reported in the figure along the wetting path hence appear to show some discrepancy due to the lack of precision of the precision calliper volume measurements, as compared to hydrostatic weighing. Under the hypothesis of saturated state, the increase in water content obtained along the wetting path corresponds to some swelling that will be further investigated later on. Conversely, up to the air entry value pressure (5 MPa) drying occurs with some shrinkage under a saturated state. The curve follows the main drying path at suction higher than 5 MPa when the sample starts desaturating. At a suction as high as 152.8 MPa, Boom clay is able to retain 5% water content. Referring to Table 1, this could indicate a high smectite content, probably closer to 33% than 17%.

Water retention data obtained by Bernier et al (1997) and Romero et al. (1999) on compacted Boom clay samples at a dry unit mass of 1.7 Mg/m$^3$ are also represented for comparison. The data show that the curve of Romero et al. (1999) is parallel, with less water being retained by the compacted sample at same suction. The curve of Bernier *et al.* (1997) is similar to that of Romero *et al.* (1999) at high suction.

Figure 4 shows the volume changes with respect to suction that correspond to the drying and wetting paths of Figure 3. A significant swelling of 18% is observed when suction is reduced to 180 kPa. A shrinkage of 15% is observed at a suction of 152.8 MPa. The slope that characterises swelling (average slope $\frac{\Delta e}{\Delta \log s} = -0.5$) is larger than the shrinkage slope (average slope $\frac{\Delta e}{\Delta \log s} = -0.1$). Bernier *et al.* (1997) found a similar trend on compacted Boom clay specimens subjected to change in suction under a small vertical load in the oedometer.

*It is difficult to provide a value of the initial suction of the sample just after extraction based on the data from the curve of Figure 3. The drying that reduced the water content from an average in-situ value close to 25% down to an average value of 21% after the storage period obviously increased the sample suction. The suction given by the figure at a water content close to the in-situ water content (25%) is equal to 400 kPa.*

As quoted by various authors including Skempton and Sowa (1963), Bishop *et al.* (1974) and Doran *et al.* (2000), there is a relation between the suction of a saturated sample and the state of stress at the depth at which the sample has been extracted. In the case of the "perfect sampling" (Bishop et al. 1974) of an isotropic elastic sample, the suction of the sample is equal to the mean effective stress supported by the sample prior to extraction. Doran et al. (2000) showed however that this was no longer true in anisotropic elastic soils even after "perfect sampling" due to coupling of volumetric strain with deviator stress during unloading. Due to the depth at which it has been sampled (223 m) and to the high corresponding in situ vertical effective stress (estimated at 2.45 MPa with an average soil unit mass $\rho = 2.1$ Mg/m$^3$ at a depth of 223 m), Boom clay can certainly be considered as a cross anisotropic material. This is also suggested by the scanning electron microscope observations carried out by Dehandschutter et al. (2004) in which bedding planes were clearly observed (note also that the Boom clay layer is gently dipping (±1°) towards the North-North-East - see Mertens et al. (2004) and Figure 1).

Based on *in-situ* stresses estimations, Horseman *et al.* (1993) calculated for Boom clay a $K_0$ value equal to 0.8. An equivalent effective mean stress ($p'$) can be derived from the vertical load ($\sigma'_v$) as follows:

$$p' = \frac{1}{3}(\sigma'_v + 2\sigma'_h) = 0.87\sigma'_v = 2.12\,MPa \qquad [1]$$

*This value is significantly higher than the suction of 400 kPa that could be estimated at w = 25% on Figure 3. The 400 kPa value seems to be too small to be explained by possible effects related to non perfect sampling and to cross anisotropy and a deeper examination of the water retention properties (Figure 3) appears to be necessary. In this regard, Figure 5 presents a zoom of the water retention curve in which an approximate estimated initial state has also been plotted by adopting the mean effective stress calculated above (2.21 MPa) as an approximate suction at a water content of 25%. Based on this hypothesis, the drying process that occurred during storage can be interpreted. The figure shows that the position of the initial point is reasonably compatible with the drying curve of the water retention curve. Note that, given the shape of the curve, this is also true for suction values down to 1 MPa. Also, the shape of the wetting path determined by the three points obtained at low suction in the oedometer illustrates an hysteresis effect typical of clays, as observed for instance by Croney (1952) on a "heavy clay soil" ($w_p$ = 26%, $w_L$ = 78%) that had been subjected to a suction cycle starting from its initial "undisturbed" state. In other words, at the same water content, the suction reached by a sample after a drying/wetting cycle can be significantly lower than the initial suction.*

*An estimation of the sample suction with a value of water content close to 21% can now be made by considering the estimated drying path of Figure 5. Accounting for the uncertainty concerning the initial state, the Figure shows that this value should be between 2*

*and 3 MPa. This range would also have been obtained by starting from an initial value of 1 MPa.*

*Oedometer tests*

The three samples used for determining the wetting path were then used for oedometer testing, with initial suction values equal to 180, 280 and 600 kPa respectively (Figure 3). The results of the oedometer compression test with suction measurement carried out with an initial suction of 280 kPa are presented as a function of time in Figure 6 as follows: (a) loading sequence (standard oedometer step loading with loading stages generally close to 24h); (b) vertical displacement; (c) pore water pressure changes, starting from an initial value of -280 kPa.

Figure 6*b* shows that each loading step induces an instantaneous settlement followed by an equilibration stage. Interestingly, instantaneous coupled peaks in positive pressure are observed at each loading step (Figure 6*c*), even when starting from an initial negative pore pressure. For vertical stresses smaller than 800 kPa, measured pore water pressures come back and stabilise at negative values. The preservation of a suction state within the sample means that no water has been expelled from the sample, resulting in a constant water content condition.

The transition from negative to positive pressure was observed when loads greater than 800 kPa were applied. Water pressure stabilised at 0 kPa after the 800 and 1600 kPa loading stages. This shows that the suction/pressure gauge worked properly in the positive pressure range, describing a standard positive pore pressure dissipation process and a decrease in water content of the sample with water expelled in the porous stone located above the sample. Similarly, instantaneous coupled pressure decreases are observed during unloading step, with suction as high as 700 kPa reached when passing from 1600 kPa to 800 kPa vertical stress. Note that this instantaneous suction value is close to the load release (800 kPa) which is reasonably compatible with saturated "inverse" consolidation.

The results of the three oedometer tests are presented in Figure 7 in diagrams giving the changes in vertical stress with respect to suction changes, once equilibrium has been reached. These diagrams show how suction is reduced by loading and then increased by unloading, starting from three different initial suctions.

During compression, a linear relation between changes in suction and vertical load is observed with slopes $ds/d\sigma_v$ that vary between -0.45 and -0.56. Compression paths also show that zero suction is reached at smaller loads when starting from smaller initial suctions, $s_i$: (a) $\sigma_{vn}$ = 1200 kPa for $s_i$ = 600 kPa; (b) $\sigma_{vn}$ = 600 kPa for $s_i$ = 280 kPa; (c) $\sigma_{vn}$ = 450 kPa for $s_i$ = 180 kPa. The slopes obtained during unloading were higher and varied from -0.59 to -0.80. The variability between the different slopes measured in Figure 7 is related to the different initial void ratios of the samples, given in Figure 4. Both samples hydrated at smaller initial suction (180 and 280 kPa) exhibited around 18% swelling with similar void ratios respectively equal to 0.86 and 0.87. For both samples, the slopes in the unloading stage (-0.74 at 180 kPa and -0.80 at 280 kPa) are larger than in the loading stage (-0.45 at 180 kPa and -0.53 at 280 kPa) with slopes slightly smaller at lower suction. The 600 kPa suction sample that exhibited a significantly smaller swelling (4%, leading to a void ratio of 0.64) has the highest value of slope during the loading phase (-0.56) and the lowest in the unloading phase (-0.59). The variability of the slopes in the loading phase (between -0.56 and -0.45) is less than in the unloading phase (between -0.59 and -0.80).

*The range of the slopes determined during the unloading sequences (between -0.59 and -0.80) gives an idea of the changes in suction that occur in Boom clay samples when releasing the vertical stress. They can be compared by the value of $s/\sigma'_v$ obtained by Bishop et al. (1974) for natural soils, in which s is the sample suction and $\sigma'_v$ the in-situ vertical*

*effective stress. Based on the sample suction obtained here (between 2 and 3 MPa), values of s/σ'$_v$ between 0.94 and 1.42 are obtained. These too high values show the effects of drying. The resulting uncertainty makes it difficult to use for a precise determination of suction changes induced by stress release.*

*Suction release under isotropic stress*

While wetting Boom clay sample at low suctions under zero stress, a swelling of around 18% due to suction release has been observed (Figure 4). This swelling is consistent with the 11-14% swelling found by Coll (2005) under a low effective pressure in the triaxial apparatus. This saturation procedure and the resulting swelling is likely to alter the initial state of the soil and to lead to unexpected low values of the yield stress, as shown by Horseman *et al.* (1993) and Sultan (1997). After Graham *et al.* (1987), during saturation under the *in-situ* state of stress, suction reduces to zero and no further swelling should occur, allowing to put the sample in contact with water and to apply a back pressure in a standard manner with no significant modification of the initial microstructure.

A test in which suction was released under a constant isotropic stress was carried out on a sample having an initial water content of 21.6% (Figure 8). In order to minimise the volume change during soaking, an isotropic stress close to the estimated initial sample suction was chosen. Based on the range of suctions estimated from the water retention curve (2 – 3 MPa), it was decided to adopt an isotropic stress of 2.5 MPa. This value is larger than the in-situ mean effective stress estimated previously (2.12 MPa). The sample was isotropically compressed from 0.1 MPa to 2.5 MPa while keeping the porous stones dry. The porous stones were then soaked and a low back pressure (50 kPa) was applied to the soil sample while measuring volume changes. The confining pressure and the back pressure were afterwards simultaneously increased in order to keep the effective stress equal to 2.5 MPa, the final back pressure being equal to 1 MPa. Finally, the soil sample was compressed up to an effective pressure of 10 MPa with a constant low pressure change rate (0.5 kPa/min). Sultan *et al.* (2002) showed that this rate was slow enough to ensure drained conditions in Boom clay. The corresponding experimental data are shown in Figure 8. Under 2.5 MPa, a slight swelling is observed during saturation ($\varepsilon_v$=1.7%), that corresponds to an increase in void ratio from 0.582 to 0.608. This swelling is much smaller than that (14%) obtained after saturation at low confining pressure by Coll (2005). Furthermore, the yield stress ($p_y$') obtained in Figure 8 is close to 5 MPa; giving an overconsolidation ratio $R_{OC}$ = 2.1, in good agreement with the data of Horseman *et al.* (1993: $R_{OC}$ = 2.4) and Coll (2005: $R_{OC}$ = 2.2). *Note however that, as commented by Horseman et al. (1987) cited in Burland (1990) based on geological evidence, the yield stress of Boom clay may be larger than the preconsolidation pressure due to mechanisms such as creep and diagenesis.*

By saturating the sample under low effective pressure Sultan (1997) obtained an underestimate value $p_y$' = 0.4 MPa. This showed that swelling may alter the natural microstructure of a swelling clay and erase the memory of the stress history by reducing significantly the overconsolidation ratio.

The data of the test of Figure 8 confirm, as suggested by Graham et al. (1987), that some precautions have to be taken before releasing the suction of a natural sample. Note however that the slight swelling observed during the test indicate that the initial suction might be higher than 2.5 MPa.

## 4. Conclusions

Suction effects were investigated in intact Boom clay block samples extracted at great depth (223 m) in the underground research laboratory of Mol (Belgium). Suction effects in deep intact samples are considered to be significant because of the relationship between suction and the in-situ stress state of the sample, as shown by various authors (Skempton 1961, Skempton and Sowa 1963, Doran et al. 2000).

Suction effects were investigated through the characterisation of the water retention and of the swelling properties of intact Boom clay. *Some drying that occurred during the storage period between the block extraction and the present experimental investigation was interpreted in terms of hysteresis effects. The sample suction was estimated to be between 2 and 3 MPa.* The swelling-shrinkage behaviour under changes in suction was investigated and an air entry value of 5 MPa was determined.

The relationship between suction and stress changes during loading and unloading sequences was also examined by running oedometer tests with suction measurements. *Slopes $ds/d\sigma_v$ between -0.59 and -0.80 were obtained in the unloading phases, to compare to the range 0.35 –0.75 of values of $s/\sigma'_v$ values proposed by Bishop et al. (1974) for natural soils. The $s/\sigma'_v$ values obtained with the block sample used in this work was in the range 0.94 – 1.42. These too high values showed the negative effects of drying.*

The effect of suction release under an isotropic stress close to the estimated suction (2.5 MPa) was finally investigated. A slight swelling (1.7%) was observed, and a further compression sequence showed that a satisfactory overconsolidation ratio (2.1) was obtained. This result confirmed the importance of taking some precautions before putting a swelling soil in contact with water prior to triaxial testing, as suggested by Graham et al. (1987).

These data confirmed the importance of suction and suction release effects, particularly in deep swelling samples of stiff clay. Obviously, some of the conclusions drawn here should be confirmed or improved by running similar tests on fresh samples just after extraction, in order to get rid of drying effects.

## 5. Acknowledgements

EURIDICE (European Underground Research Infrastructure for Disposal of nuclear waste In Clay Environment, Mol, Belgium) is gratefully acknowledged for funding the work presented in this paper. This work is part of the PhD thesis prepared at ENPC Paris by the first author. The financial support of ENPC is also acknowledged. The Authors are also grateful to the Referee and Assessor whose comments greatly helped in improving the paper.

# List of tables

# List of figures

## List of Notations

$d$      sample diameter

$e$      void ratio

$h$      sample height

$K_0$      coefficient of earth pressure at rest

$p_c'$      preconsolidation pressure

$R_{OC}$      overconsolidation ratio

$s$      suction

$s_i$      initial suction

$S_r$      degree of saturation

$w$      water content

$w_i$      initial water content

$\varepsilon_V$      volumetric deformation

$\sigma_h$      horizontal stress

$\sigma_v$      vertical stress

$\sigma'_v$      vertical effective stress

$\sigma_{vn}$      vertical stress when suction decreases to 0

|  | Decleer et al. (1983) (%) | Al-Mukhtar et al. (1996) (%) | SAFIR 2 (2001) (%) |
|---|---|---|---|
| ▪ Clay minerals | 50 | 62 | 52 |
|     Illite | 12 | 16 | 28 |
|     Kaolinite | 5 | 13 | 6 |
|     Smectite | 33 | 33 | 17 |
|     Chlorite |  |  | 3 |
|     Glauconite |  |  | 3 |
| ▪ Quartz | 35 | 20-25 | 20 |
| ▪ Calcite, Dolomite | 1 |  | 1-5 |
| ▪ Pyrite | 1 | 4-5 | 1-5 |
| ▪ Feldspar |  |  | 5-10 |
|     Microcline | 9 | 4-5 |  |
|     Plagioclase | 4 | 4-5 |  |
| ▪ Organic material |  |  | 1-3 |

**Table 1. Mineralogical composition of Boom clay according to Decleer et al. (1983), Al Mukhtar et al. (1996) and SAFIR 2 (2001).**

| Boom clay | Belanteur et al. (1997) | Dehandschutter et al. (2005) |
|---|---|---|
| Unit mass of solid (Mg/m$^3$) | 2.67 |  |
| Unit mass (Mg/m$^3$) |  | 1.9 |
| Liquid limit $w_L$ | 59-76 | 70 |
| Plastic limit $w_P$ | 22-26 | 25 |
| Plastic index $I_P$ | 37-50 | 45 |
| Water content (%) |  | 25-30 |
| Natural porosity (%) |  | 35 |
| Poisson's ratio |  | 0.4 |
| Internal friction angle (°) |  | 18 |
| Permeability (m/s) |  | $10^{-12}$ |

**Table 2. Geotechnical properties of Boom clay according to Belanteur et al. (1997) and Dehandschutter et al. (2005).**

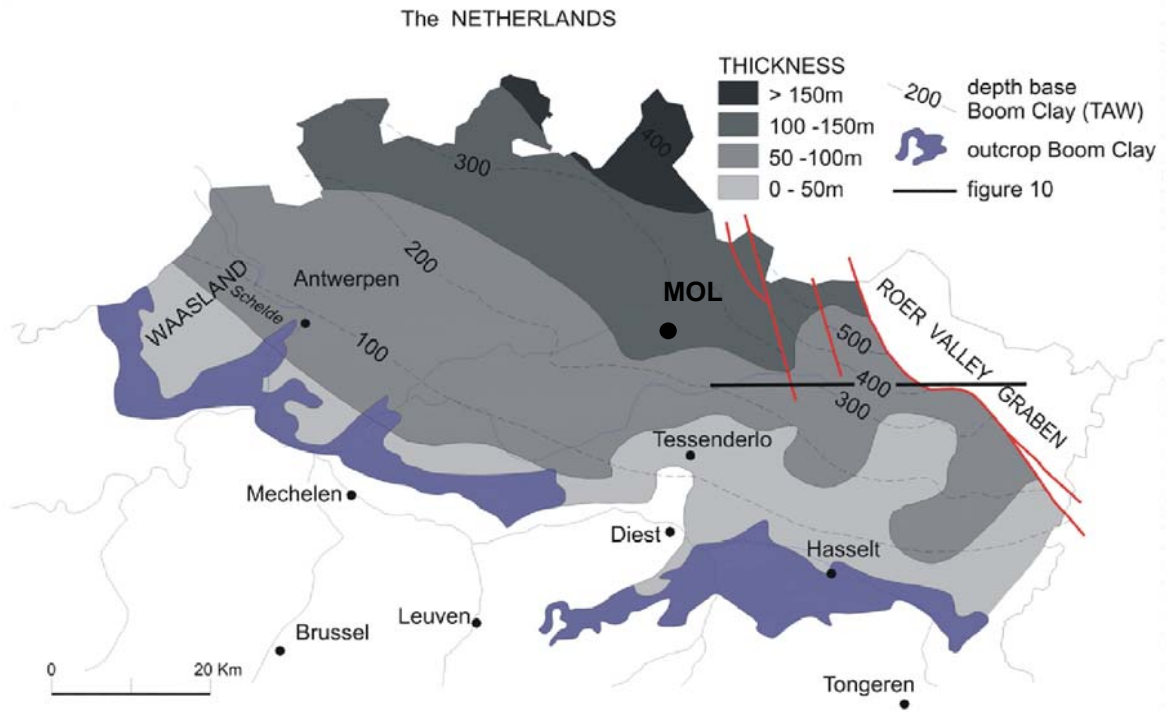

**Figure 1.** Boom clay formation in the North of Belgium. The Underground Research Laboratory is located near the city of Mol. The clay formation is gently dipping (±1°) towards the North North-East (Mertens *et al.*, 2003).

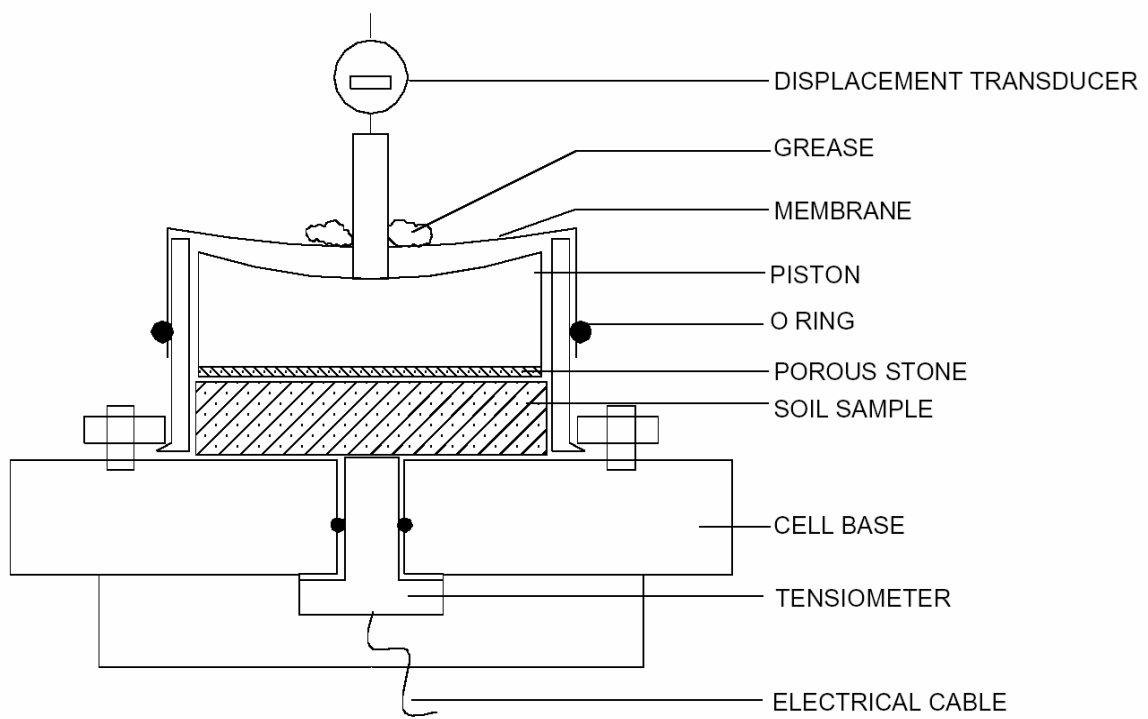

**Figure 2.** Oedometer cell equipped with a tensiometer at the basis.

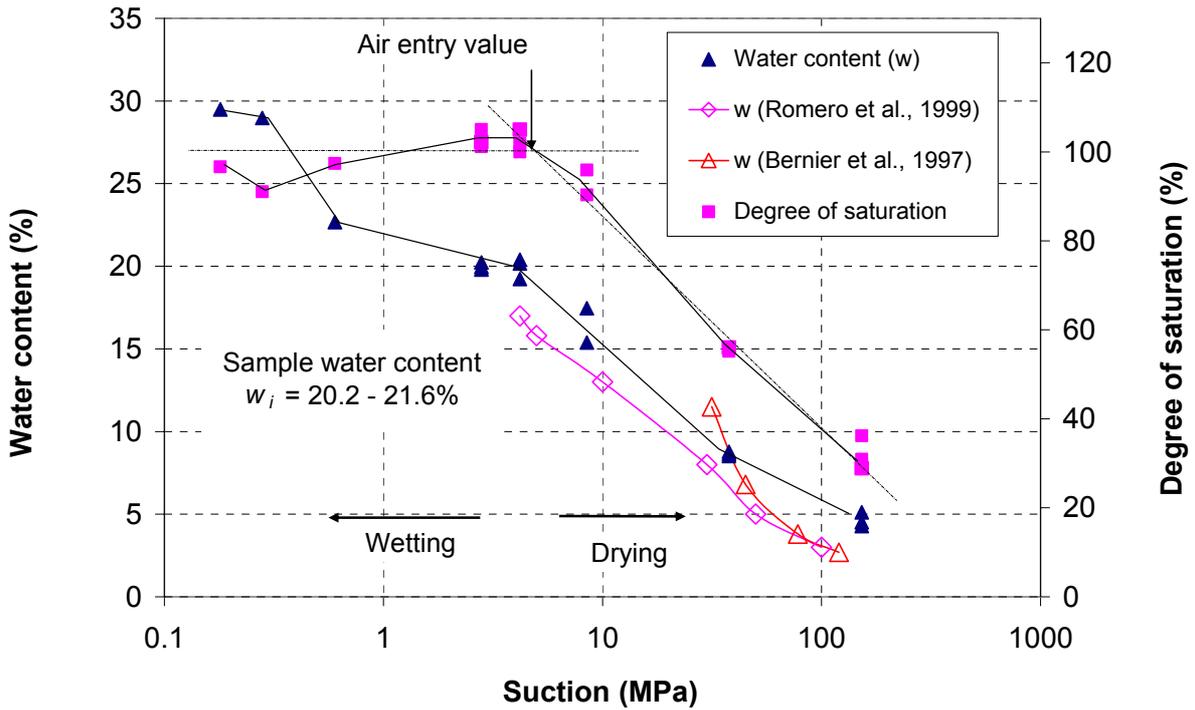

**Figure 3.** Water retention properties of Boom clay sample in terms of changes in water content and degree of saturation versus suction (wetting and drying paths followed respectively starting from a 21% initial water content).

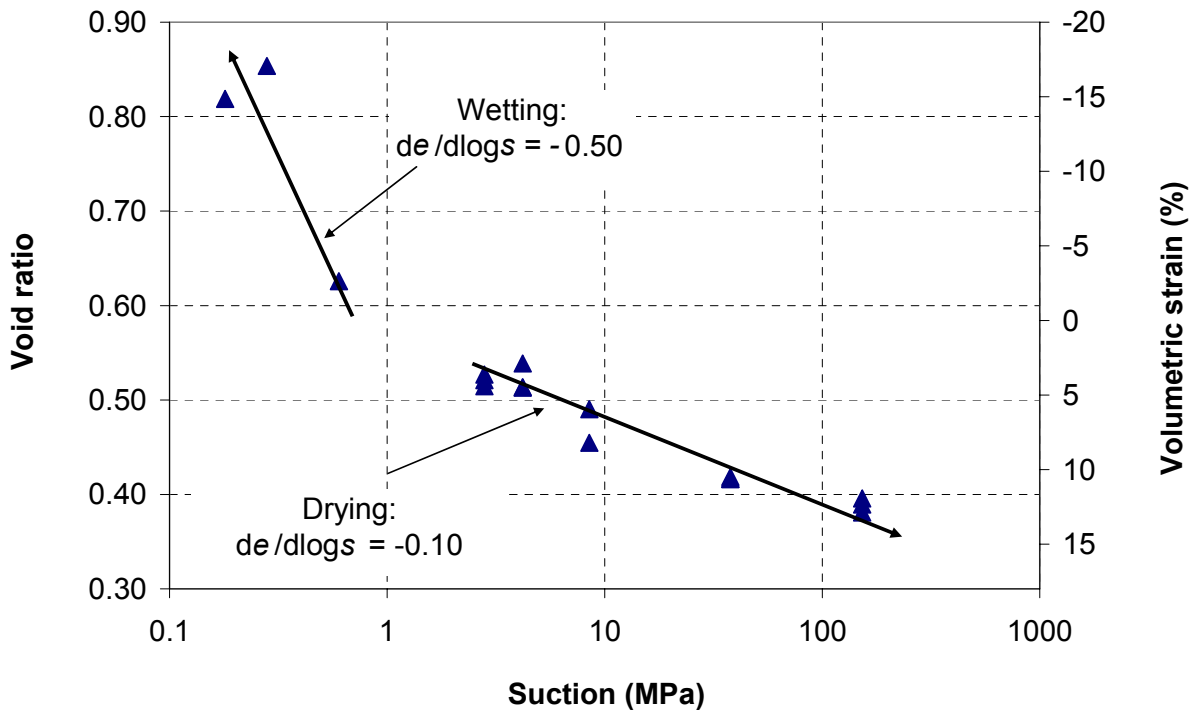

**Figure 4.** Volume changes of Boom clay samples under suction changes during drying and wetting.

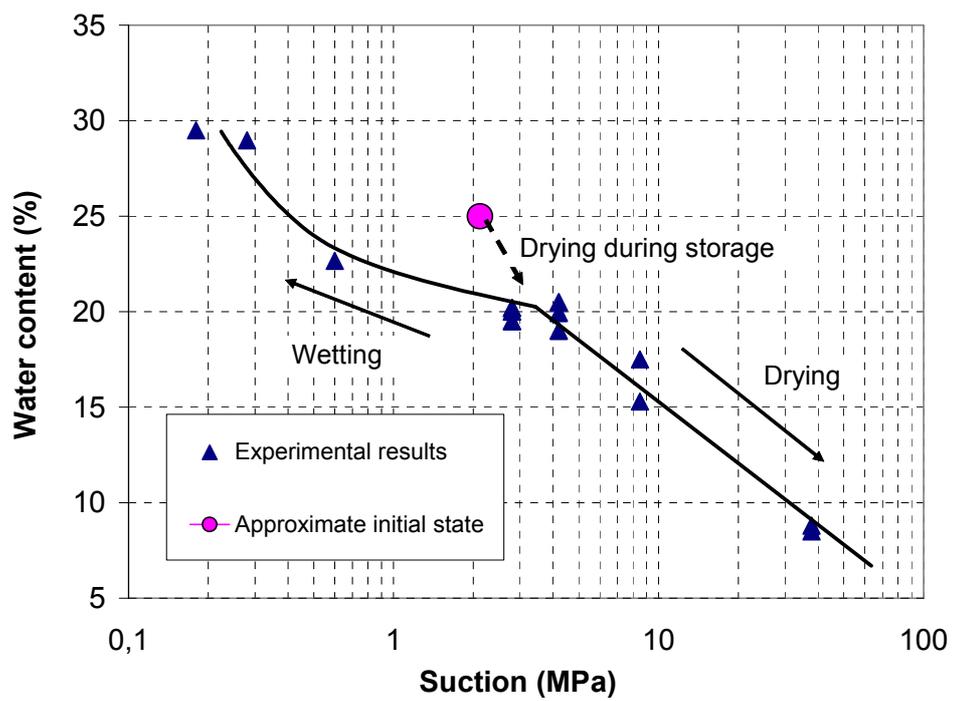

**Figure 5.** Hysteresis effects in the lower suction range

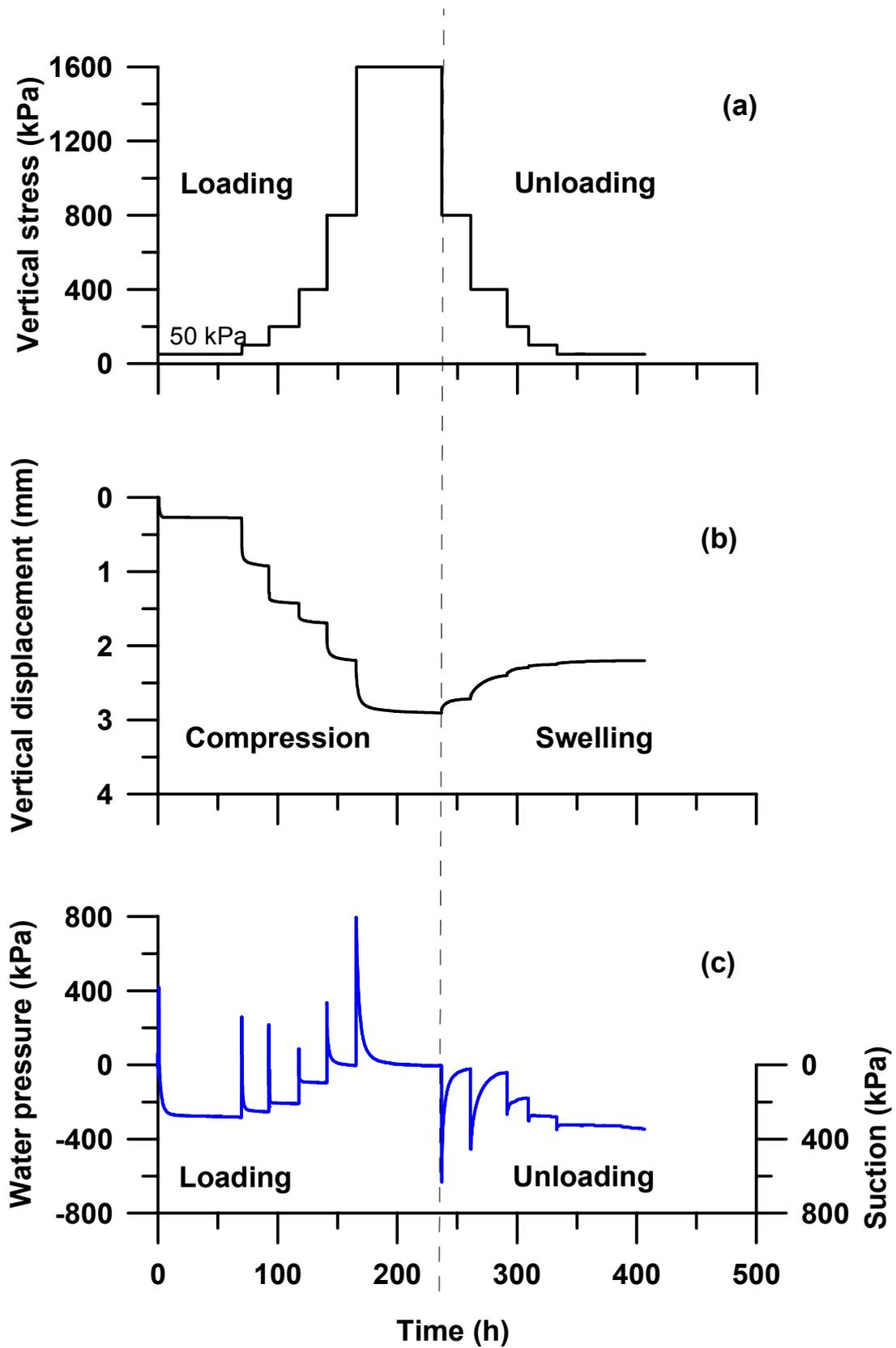

**Figure 6.** Oedometer compression test with suction measurement. Vertical stress (a), vertical displacement (b) and suction changes (c) are given as a function of elapsed time (initial suction 280 kPa).

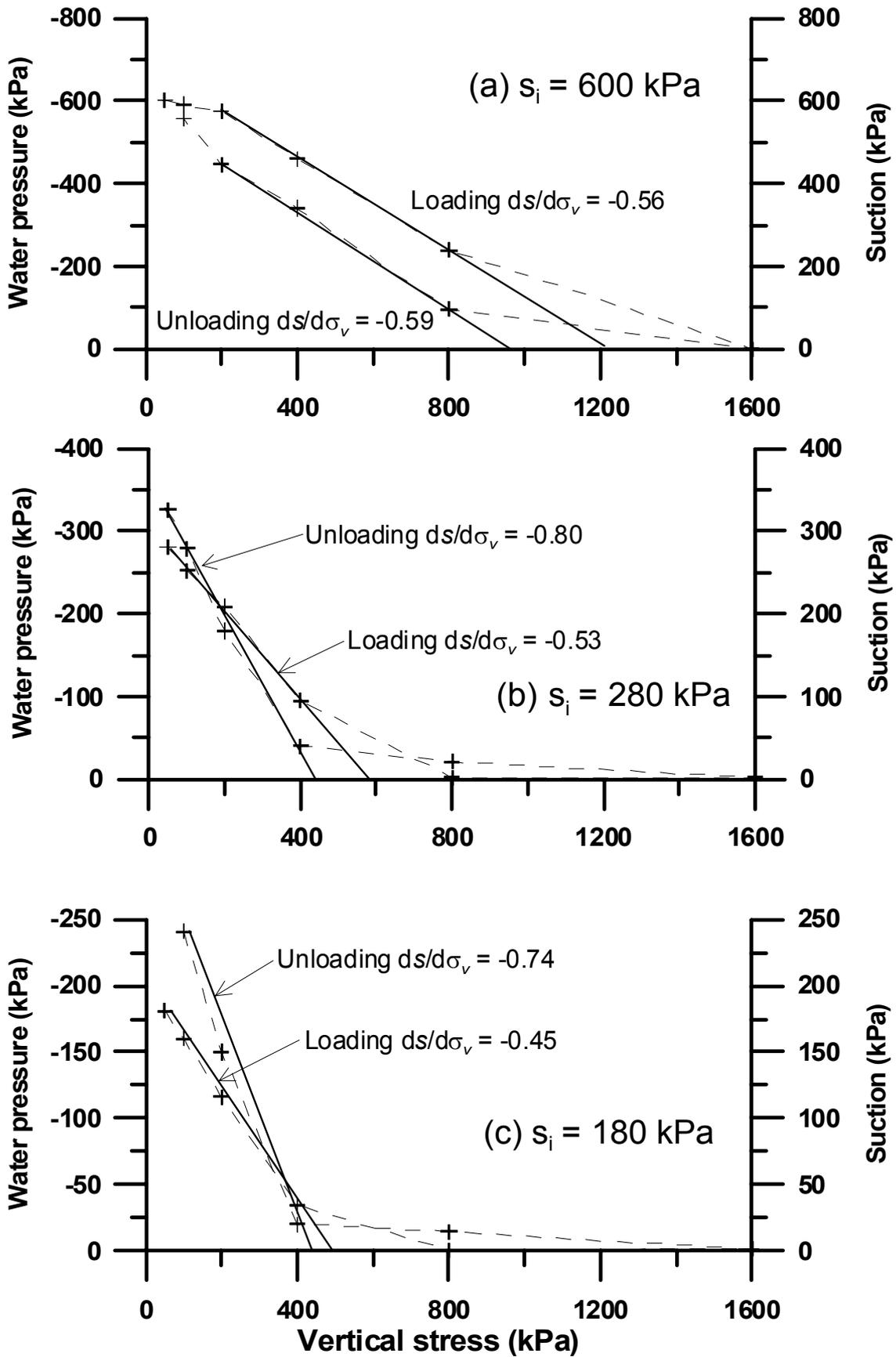

Figure 7. Suction variation under vertical stress change during oedometer compression for three tests with initial suction equal to 600, 280 and 180 kPa respectively.

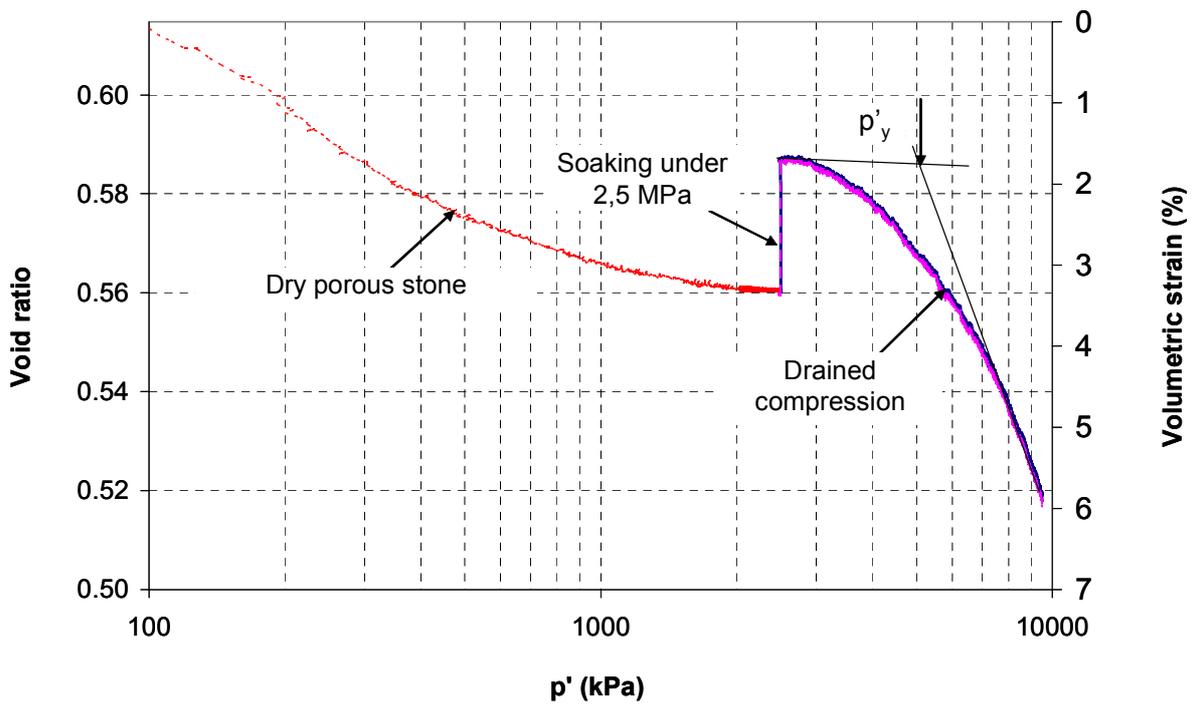

Figure 8. Isotropic compression test.